\documentclass[aps,prl,preprint,groupedaddress,longbibliography,floatfix]{revtex4-2}
\usepackage{graphicx,amsmath,amssymb}
\graphicspath{{./Figures/}}

\usepackage[percent]{overpic}
\usepackage{color}
\usepackage[normalem]{ulem}
\usepackage[utf8]{inputenc}
\usepackage{mathtools}
\newcommand{\sjt}[1]{{\color{black}#1}}

\begin{document}


\title{Non-equilibrium thermodynamics in driven macroscopic self-assembly}


\author{S. J. Thomson$^{1, 2}$}
\email{Corresponding author: stuart.thomson@bristol.ac.uk} 
\author{J.-W. Barotta$^2$, and D. M. Harris$^2$}

\affiliation{$^1$School of Engineering Mathematics and Technology, University of Bristol, Ada Lovelace Building, University Walk, Bristol, BS8 1TW}
\affiliation{$^2$School of Engineering, Center for Fluid Mechanics, Brown University, 184 Hope Street, Providence, RI 02912}


\date{\today}

\maketitle

\textbf{Equilibrium statistical mechanics provides a robust framework for characterizing phase transitions in systems whose microsopic dynamics are time-reversible. Efforts to develop and validate theoretical frameworks for time-irreversible, non-equilibrium systems are constrained by experimental data that capture only partial measurements of the system dynamics. We herein overcome this limitation using a tunable macroscopic platform for non-equilibrium physics: millimetric spheres bound by capillary attractions at the fluid interface and driven out of equilibrium by a field of supercritical Faraday waves. The external driving induces correlated fluctuations in the particle trajectories, which in turn excite structural rearrangements between distinct metastable cluster topologies. By tracking all microstate transitions experimentally, we directly measure a non-zero entropy production rate reflecting broken detailed balance and quantifying the system's departure from equilibrium. The measured stochastic dynamics are in quantitative agreement with a many-body active Ornstein-Uhlenbeck model, thus establishing a bridge to a wider class of athermal, self-propelled systems at the microscale. These results invite parallel studies of non-equilibrium self-assembly kinetics using active colloids or passive particles immersed in bacterial baths whose dynamics and irreversibility are likewise governed by correlated active forces and tunable inter-particle interactions.}

Non-equilibrium processes abound in nature and technology from the development and self-organization of living matter \cite{grzybowski2009self, tan2022odd} to the directed assembly of colloids \cite{chakraborty2022self, tang2022control, bishop2023active} and the coordinated behaviour of macroscopic robot collectives \cite{slavkov2018morphogenesis, yu2023programmable}. An important aspect of these and myriad other systems in physics and biology are phase transitions, wherein otherwise identical constituents of the system can occupy, and transition between, different microstate configurations. As a consequence of time-reversal symmetry, in equilibrium statistical mechanics these transitions follow the directions of steepest ascent and descent in a free-energy landscape \cite{vanden2010transition, o2022time}. For non-equilibrium systems whose dynamics break time-reversal symmetry, this framework no longer applies and hence attention has turned to developing a corresponding theory for phase transitions out of equilibrium \cite{zakine2023minimum, heller2024evaluation}. However many microscopic or biological systems that have inspired the development of such frameworks are limited to partial experimental observations \cite{martinez2019inferring, skinner2021improved} or contain hidden degrees of freedom that are not readily measured, rendering experimental validation difficult and often limited to comparisons with idealized theoretical models. Moreover, for non-equilibrium physics to realize its technological potential, it is important to determine to what extent synthetic, non-living systems can exhibit dynamical and statistical signatures akin to their animate counterparts. 
\begin{figure}[htbp]
\includegraphics[width=\columnwidth]{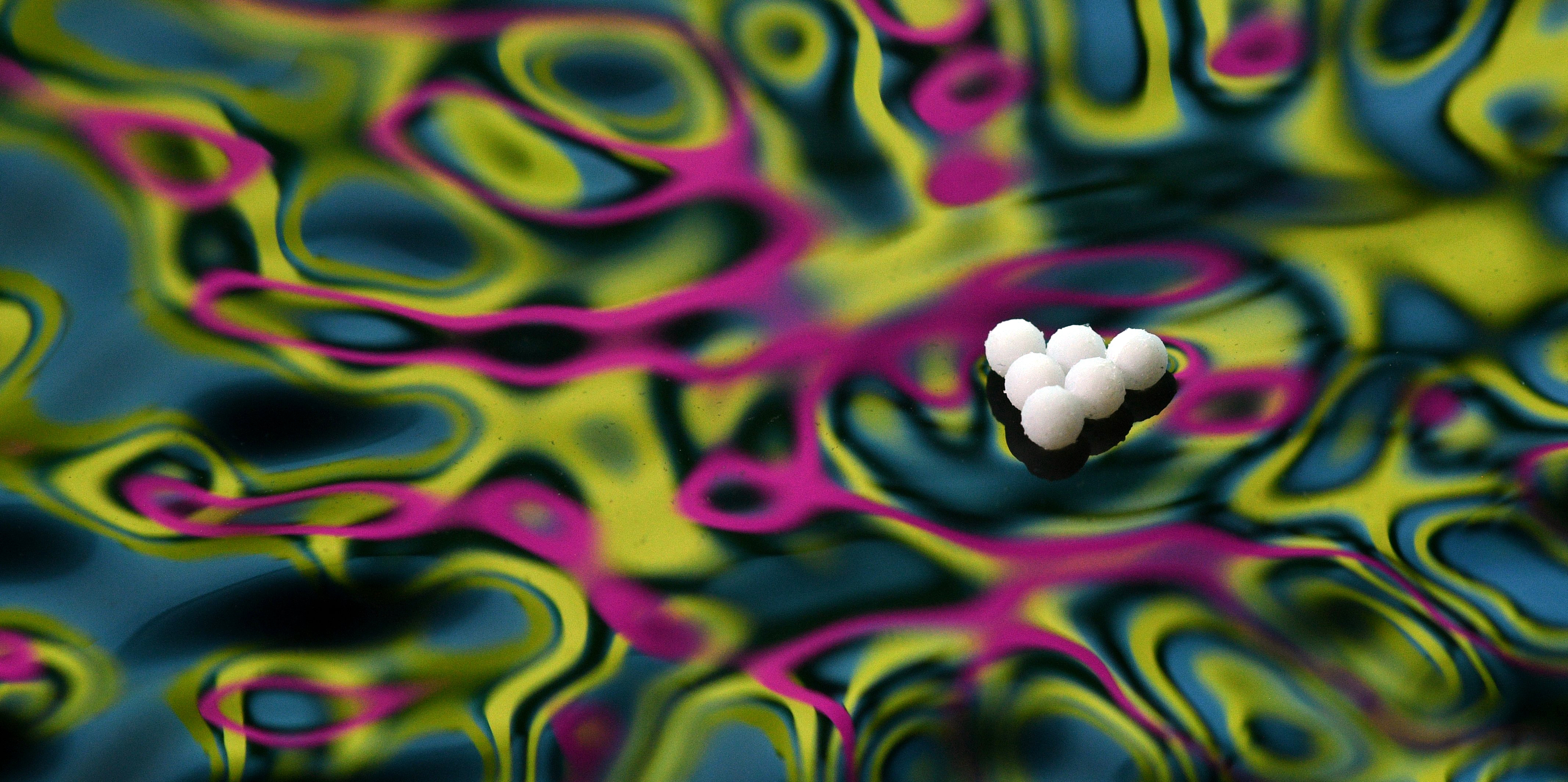}
\caption{Oblique perspective of a cluster of six spherical particles, each of radius 0.079 centimeters, momentarily occupying a triangle configuration in a field of supercritical Faraday waves of characteristic wavelength $\lambda = 0.84$ centimeters. The Faraday waves are visualized using a reflected color pattern \cite{harris2017visualization}. The complex fluid environment due to the Faraday waves and horizontal turbulent surface flows plays the role of an active bath.}
\label{fig:Fig1}
\end{figure} 

To overcome these limitations, we herein present a macroscopic experimental platform for non-equilibrium physics consisting of few-particle clusters of spherical particles bound by capillary attractions at the fluid interface, driven out of equilibrium by a field of supercritical Faraday waves that serve as a tunable \emph{active bath} (Fig.\ \ref{fig:Fig1}). This macroscopic realization offers full access access to particle dynamics and transitions between microstates excited by the external driving. We leverage this accessibility to first demonstrate how the cluster occupation probabilities and transition statistics depend on the strength of the vibrational forcing. \sjt{We then use these quantities to compute the entropy-production rate of the system, which is non-zero and increases monotonically with the forcing, reflecting broken detailed balance and providing a clear signature of the system’s departure from equilibrium. The stochastic dynamics and emergent statistical observables of the system are in quantitative agreement with a many-body active Ornstein-Uhlenbeck (AOU) model whose nonequilibrium behaviour derives from an imbalance between finite-time correlations in active noise and dissipation}. \sjt{The AOU model has been shown elsewhere to successfully describe the dynamics of passive tracers in bacterial baths \cite{maggi2014generalized, fodor2016howfar} while correlated active forces are a hallmark of a wider class of non-equilibrium self-propelled systems. By demonstrating that entropy production and broken detailed balance manifest universally across scales, our results establish a quantitative, experimental link between the stochastic thermodynamics of macroscopic systems and microscopic active matter, motivating corresponding investigations of phase transitions in systems composed of active colloids or immersed in other athermal environments}.



\begin{figure*}[htbp]
\includegraphics[width=\textwidth]{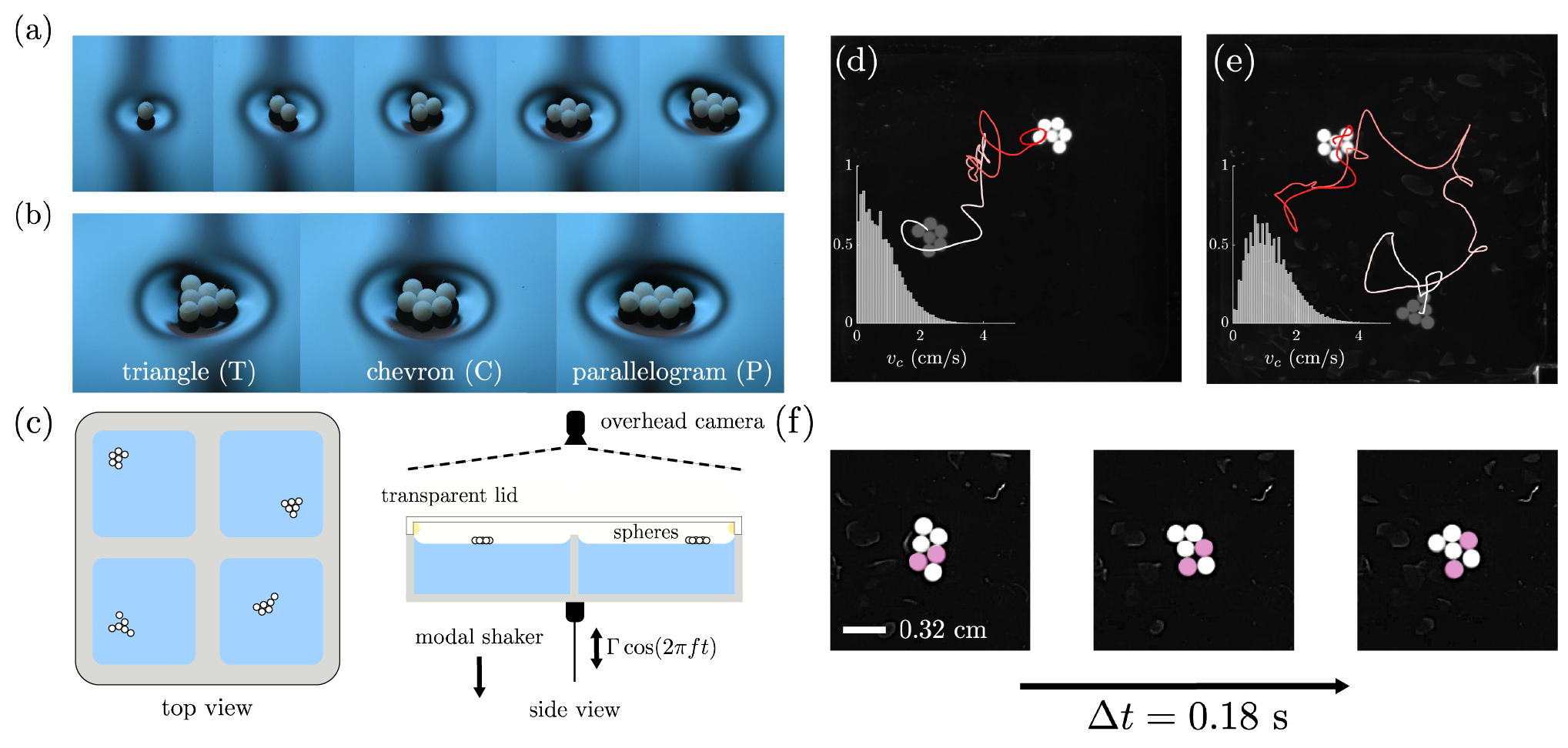}
\caption{Faraday-wave driven structural rearrangements of capillary-bound six-particle clusters. (a) Oblique images of the unique ground-state configurations formed when fewer than six spherical particles of radius 0.079 centimeters (cm) are placed on the quiescent surface of a water-glycerol mixture. (b) Addition of a sixth particle forms one of three metastable configurations: triangle (T), chevron (C, ground state), or parallelogram (P). (c) Schematic of our experimental set-up. Four clusters are confined to square-shaped corrals filled with a water-glycerol mixture. The entire set-up is vibrated vertically with acceleration $\Gamma$ and frequency $f$ above the Faraday threshold $\Gamma_F = 0.9g$. Typical experimental cluster trajectories (solid lines) over a 15 second interval are shown for (d) $\Gamma = 1.0g$ and (e) $\Gamma = 1.2g$. The white-to-red colormap along the cluster trajectory history indicates increasing time. The inset of both (d) and (e) plots the long-time experimental distribution of $v_c = \lVert\sum_{i = 1}^6 \dot{\mathbf{x}}_i(t)\rVert_2$, the instantaneous speed of the cluster center of mass. The mode of the distribution is 0.83 cm/s for $\Gamma = 1.0g$ and 1.16 cm/s for $\Gamma = 1.2g$ showing a speed enhancement of the particles as $\Gamma$ is increased. (f) Example of a P-to-C transition when a single bond between two particles (shaded in pink) breaks. A typical transition occurs over a sub-second timescale (0.18 s in this example).}
\label{fig:Fig2}
\end{figure*} 
When placed on the quiescent surface of a water-glycerol mixture (8:2 by volume, density $\rho = 1058.1$ kg/m$^3$, surface tension $\sigma = 0.0718$ kg/s$^2$ and \sjt{dynamic viscosity 0.00189 kg/m$\cdot$s} \cite{volk2018density}), fewer than six spherical Polytetrafluoroethylene (PTFE) particles (radius $R = 0.79\times 10^{-3}$ m) coated with a commercially available spray (Rust-Oleum NeverWet) self-assemble into a unique ground-state configuration through the equilibration of capillary and contact forces, a phenomenon colloquially known as the ``Cheerios effect'' (Fig.\ \ref{fig:Fig2}(a), Methods, Supplementary Section 1.1) \cite{vella2005cheerios,ho2019direct, protiere2023particle}. The addition of a sixth particle gives rise to three geometrically distinct, \emph{metastable} cluster configurations, referred to herein as chevron (C), triangle (T), and parallelogram (P) (Fig.\ \ref{fig:Fig2}(b)). In the limit of small interfacial deformations and small Bond number, the capillary interaction-energy between six spherical particles floating at the fluid interface may be approximated by $U = -U_0\sum_{i = 1}^{6}\sum_{j > i} K_0(|\mathbf{x}_i - \mathbf{x}_j|/l_c)$ where $K_0$ is the zeroth-order modified Bessel function of the second kind, $\mathbf{x}_i$ denotes the horizontal position of the $i$th particle in the cluster, the capillary length of the fluid $l_c = \sqrt{\sigma/\rho g} = 2.64\times 10^{-3}$ m and $g = 9.81$ m/s$^2$ is acceleration due to gravity \cite{kralchevsky2000capillary, vella2005cheerios}. The energy parameter $U_0 = 2\pi\sigma R^2 \text{Bo}^2 \Sigma^2 \approx 4.5$ nanojoules (nJ) where the Bond number $\text{Bo} = (R/l_c)^2 = 0.1 \ll 1$ and $\Sigma = 1.38$ is the dimensionless resultant weight of the particle (Supplementary Section 2.2.2)    \cite{vella2005cheerios}. Hence $U\rightarrow 0$ for particle spacings much greater than $l_c$. Computing $U$ for the C, T, and P configurations we find $U^{\text{C}} = -41.3$ nJ, $U^{\text{T}} = -40.9$ nJ and $U^{\text{P}} = -40.4$ nJ and hence the C configuration is the true ground state of the six-particle system. (The hypothetical limit $l_c/R\rightarrow 0$ corresponds to short-range nearest-neighbor coupling in which case the C, T and P configurations are found to be energetically degenerate akin to colloidal clusters \citep{perry2015two}.) 
\begin{figure*}[htbp]
\centering
\includegraphics[width=\textwidth]{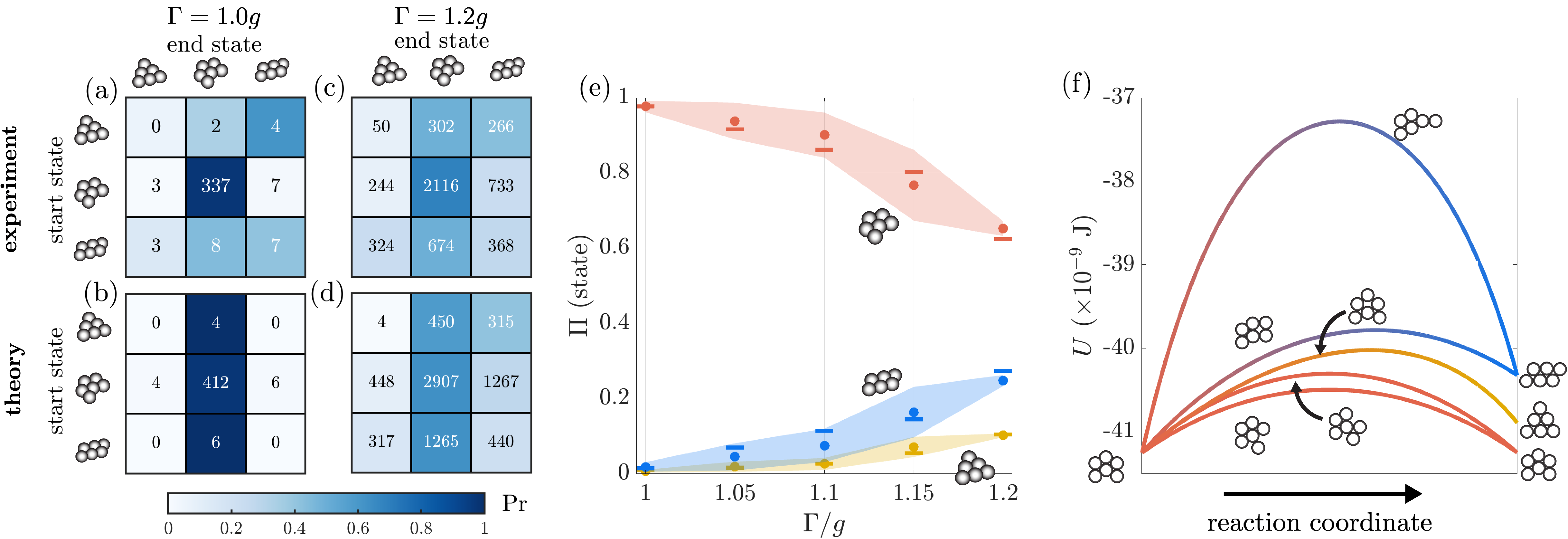}
\caption{Control of six-particle assembly and transition statistics with vibrational forcing. Transition probability matrices for (a)--(b) $\Gamma = 1g$ and (c)--(d) $\Gamma = 1.2g$. \sjt{The top figure in each pair are the experimental results, the bottom figure the corresponding theoretical prediction.} The colors in (a)--(d) correspond to the conditional probability Pr(end$\lvert$start) of finishing in a particular end state following a transition from one of three starting states. Numerical values of the mean number of pairwise transitions taken over the four quadrants of the experiment are overlain in (a) and (c), with the corresponding theoretical counts overlain in (b) and (d). Transitions that terminate in a C state have higher probability regardless of start state. (e) Experimental statistical steady-state occupation probabilities for six-particle clusters for $\Gamma/g = (1, 1.05, \ldots, 1.2)$ \sjt{where the points represent the mean $\Pi$ computed over the four quadrants of the experiment. The shaded regions represent the standard error around the mean}. Theoretical occupation probabilities predicted from \sjt{simulations of Eqns.\ \eqref{eqn:AOUP}} are plotted as a function of the \sjt{corresponding values of $D$ and $\tau$ determined from the mean-squared displacement and velocity-correlation function (solid dash)}. (f) Plots of the variation in capillary energy, $U$, when transitioning from a C state through the breaking of a single bond. The lowest activation energies belong to C-C transitions where one or two particles move around the periphery of the cluster in a hinge-like motion, illuminating why the C-C transitions are favored.}
\label{fig:Fig3}
\end{figure*}
In the absence of external forcing, the capillary forces are strong enough that the passive, non-Brownian particles used in the present study are irreversibly locked-in to one of the C, T, or P cluster configurations \cite{vassileva2007fragmentation, liu2018capillary, lagarde2020probing}. This lack of dynamic reconfigurability contrasts \sjt{with scenarios where} thermal fluctuations are sufficient to \sjt{overcome inter-particle forces} and excite transitions between microstates \cite{perry2012real,perry2015two}. However, agitation of the particles by Faraday waves and accompanying turbulent surface flows \cite{sanli2014collective, welch2014ballistic, xia2019tunable, francois2020nonequilibrium} causes the clusters to translate across the fluid surface and drives structural rearrangements between the C, T, and P cluster topologies (Figs.\ \ref{fig:Fig2}(c)--(f) and \sjt{Supplementary Video 1}). To track the particle dynamics, we confined four sets of six-particle clusters to four separate but identical illuminated square chambers, each filled with $9.0\pm 0.1$ milliliters of water-glycerol solution (Fig.\ \ref{fig:Fig2}(c), Methods, Supplementary Section 1.2). A small concave meniscus on the perimeter of each chamber repelled the clusters away from the boundary. The entire set-up is mounted on an in-house electromagnetic shaker vibrating vertically with acceleration $\Gamma\cos(2\pi f t) $ where $\Gamma$ is the maximum acceleration, $f$ is the vibrational frequency, and $t$ is time. An external air-bearing guides the motion along a single axis and a closed-loop feedback system ensures that $\Gamma$ remains within a tolerance of $\pm 0.001g$ \cite{harris2015generating}. For all our experiments $f = 60$ Hz  and for the fluid parameters and corral geometry reported here, the Faraday threshold (the critical value of $\Gamma$ for which supercritical Faraday waves spontaneously appear on the fluid surface) of the fluid is $\Gamma_F = 0.90 \pm 0.03 g$. 
(\sjt{We also tested the experiment at $f = 80$ Hz and found similar phenomenology but increasing $f$ beyond 80 Hz increased the tendency of clusters to disintegrate, significantly inhibiting the collection of reliable statistical data of the particle dynamics}.)
For each experimental run, the particles are filmed for 1 hour and 6 minutes using a camera recording at 40 frames-per-second mounted directly over the bath surface, capturing thousands of transitions over the lifetime of the experiment (Supplementary Section 1.3). 

 Cluster configurations, and hence transitions between the C, T, and P microstates, are identified from each cluster's adjacency matrix in every frame of post-processed particle-tracking data (Supplementary Section 1.4) \cite{perry2015two}. A typical cluster rearrangement involves a single bond between two neighbouring particles breaking before the cluster settles into a new state (Fig.\ \ref{fig:Fig2}(f) and \sjt{Supplementary Video 2}), similar to thermally driven colloids \cite{perry2015two}. As $\Gamma$ is increased there is an accompanying speed enhancement of the particles (inset Figs.\ \ref{fig:Fig2}(d) and (e)) and rearrangements can involve the breaking of multiple bonds, the particles taking a more circuitous route between states thereafter \sjt{(see Supplementary Video 2)}. On occasion a particle can detach from the cluster altogether and freely translate around the bath surface before eventually docking onto the cluster once more under the action of attractive capillary forces. 

The experimental transition count matrices and resulting transition probability matrices for $\Gamma = 1g$ and $\Gamma = 1.2g$ are shown in Figs.\ \ref{fig:Fig3}(a) and (c), respectively. For $\Gamma = 1g$, \sjt{the particles spend the majority of time in the C configuration (the ground state)} and most transitions occur between two chevron states (transitions between chevron chiral enantiomers) that involves one or two particles moving around the perimeter of the cluster in a hinge-like motion. As $\Gamma$ is increased \sjt{there is a corresponding increase in the total number of transitions} and \sjt{the number of transitions that occur between any two states, yet transition pathways terminating in a C state are favoured regardless of whether the cluster starts in a C, P or T state}. The appreciable number of T-T transitions for higher $\Gamma$ is due to the aforementioned detachment events where one or two particles leave the cluster only to later rejoin. 

\sjt{The steady-state occupation probabilities (or stationary distribution) $\Pi$ of the C, T and P cluster configurations are the left eigenvectors of the transition probability matrix $P$, specifically the solutions of $\Pi^T P = \Pi^T$.} Across the range of $\Gamma$ considered, we find that C states are more propitious followed by P and then T (Fig.\ \ref{fig:Fig3}(e)). When $\Gamma = 1g$ there are relatively few instances of T and P configurations compared to C. However, as $\Gamma$ is increased incrementally to $\Gamma = 1.2g$ \sjt{in steps of 0.05$g$}, there is a monotonic reduction in $\Pi$ (chevron) and a corresponding increase in the yield of both T and P configurations. \sjt{The foregoing observations highlight the role that fluctuation amplitude can play in facilitating exploration of complex energy landscapes in a more generic class of interacting active particle systems with potentially numerous microstates.} 

In the present capillary system the range of particle-particle interactions is set by the capillary length of the fluid which in turn is comparable to the particle size. Consequently, $U$ varies continuously along the entire reaction coordinate as the particles traverse transition pathways. To gain further insight into why C states are favored, we consider the one-dimensional energy landscape formed by a subset of possible transition paths wherein a single bond breaks in a C cluster and the particles go on to form a C, P or T configuration (Fig.\ \ref{fig:Fig3}(d)). In addition to the C configuration being the true ground state of the system, we find that lower activation energies are required to excite hinge-like C-C transitions than any other pathway and hence these pathways will be most favored. Beyond energetic considerations, symmetry also plays a role: for example, C-P transitions are more favored than C-T transitions despite having slightly higher activation energy since there are 4 ways to enter into a C-P transition and only 1 for C-T transitions in this simplified landscape. \sjt{This is reflected in the transition/count matrices (Fig.\ \ref{fig:Fig3}(a) and (c)) where there are roughly 3 times as many C-P transitions as C-T.} 

\begin{figure*}[htbp]
\centering
\includegraphics[width=\textwidth]{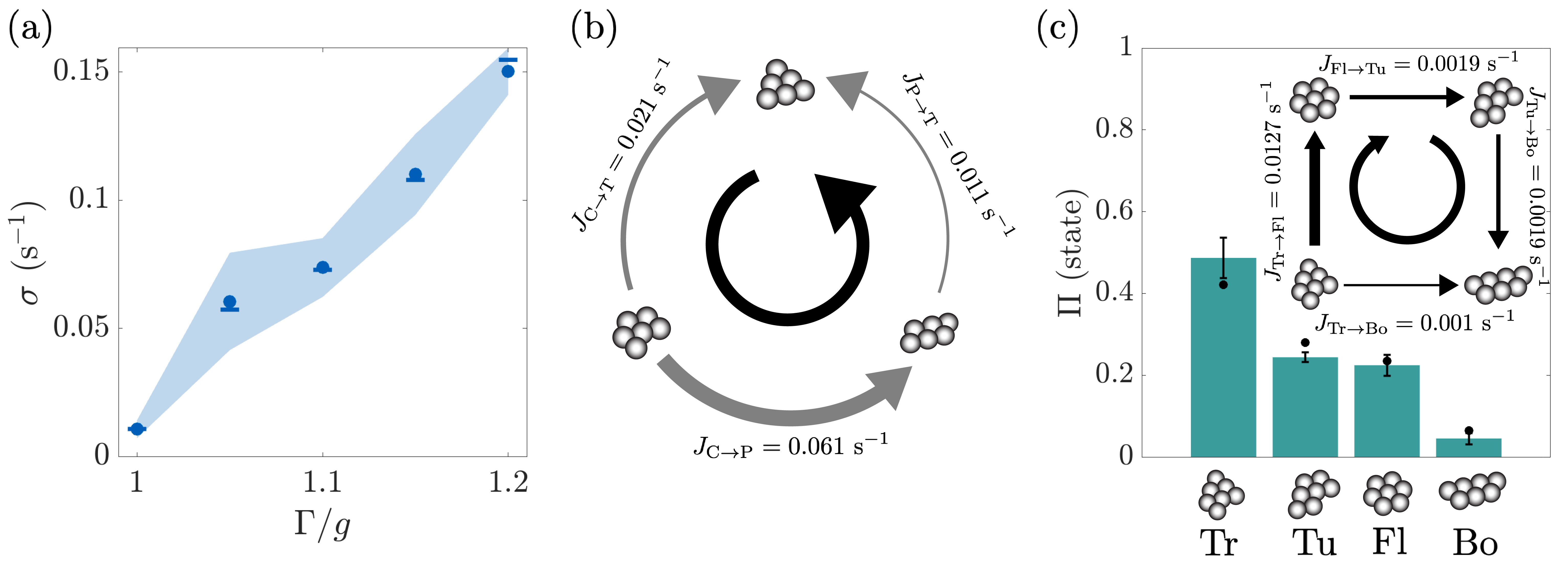}
\caption{Nonequilibrium thermodynamics and broken detailed balance at the macroscale. (a) The mean experimental entropy production rate $\sigma$ as a function of the vibrational forcing $\Gamma$ (dots) compared to theoretical predictions from Eqns.\ \eqref{eqn:AOUP} (solid lines) for $N = 6$ particles. The mean is taken over the four quadrants of the experiment. The parameters $D$ and $\tau$ for the theoretical model are determined from the mean-squared displacement $\sum_{j = 1}^N \langle |\mathbf{x}_j(t) - \mathbf{x}_j(0)|^2 \rangle/N$ and velocity-correlation function $\sum_{j = 1}^N\langle\dot{\mathbf{x}}_j(t)\cdot\dot{\mathbf{x}}_j(t')\rangle/N$, where $\langle\cdot\rangle$ denotes the average over particle trajectories. The shaded region represents the standard error taken over each quadrant of the experiment. (b) Steady-state probability current diagram for the representative case $N = 6$ and $\Gamma = 1.2g$, illustrating broken detailed balance via a persistent flux loop in the direction $\text{T}\rightarrow \text{C}\rightarrow \text{P}$. The quantities $J_{i\rightarrow j} = \Pi_i Q_{ij} - \Pi_j Q_{ji}$ are the net probability currents between microstates $i$ and $j$. (c) Experimental stationary distribution for a 7-particle experiment at $\Gamma = 1.2g$ (green bars) with theoretical predictions of the model \eqref{eqn:AOUP} shown by the (black) dots. The error bars on top of each probability bar signifies one standard deviation on the mean stationary distribution taken over the four quadrants of the experiment. Inset: steady-state probability current diagram illustrating broken detailed balance via one of the directed cycles $\text{Fl}\rightarrow\text{Tu}\rightarrow\text{Bo}\rightarrow\text{Tr}$.}
\label{fig:Fig4_EPR}
\end{figure*}

\sjt{Although equilibrium arguments provide some intuition, closer inspection of the transition statistics of the present system reveals that the system dynamics are not readily understood within an equilibrium framework. Equilibrium statistical mechanics predicts that $\Pi$ is Boltzmann distributed with $\Pi$(C), $\Pi$(P) $\rightarrow 2/5$ and $\Pi$(T) $\rightarrow 1/5$ for $U\ll \beta^{-1}$ and $\Pi$(C) $\rightarrow 1$ for $U\gg \beta^{-1}$, accompanied by a crossover in $\Pi$(T) and $\Pi$(P) as the inverse Boltzmann energy $\beta$ varies (Supplementary Section 2.1, Supplementary Fig.\ 1). However, no such crossover is observed in our data (Fig.\ \ref{fig:Fig3}(e)). This discrepancy motivates us to consider the canonical measure of the departure of a stochastic process from equilibrium, namely the entropy production rate (EPR) $\sigma = \sum_{i\neq j} \Pi_i Q_{ij}\log(\Pi_i Q_{ij}/\Pi_j Q_{ji})$ where $Q_{ij}$ is the transition rate at which the system transitions from state $i$ to $j$ \cite{seifert2012stochastic}. For an equilibrium process, forward transitions and their reverse counterparts are equally likely to be observed and the process is said to satisfy detailed balance. This corresponds to $\Pi_i Q_{ij} = \Pi_j Q_{ji}$ for all $i$ and $j$, or $\sigma\equiv 0$. Applying this measure to our experimental data, we instead find a non-zero and monotonically increasing EPR with $\Gamma$ (Fig.\ \ref{fig:Fig4_EPR}(a)). By computing the net probability currents $J_{i\rightarrow j} = \Pi_i Q_{ij} - \Pi_j Q_{ji}$ between each microstate of the system, we find that this irreversibility originates from an imbalance in cyclic probability currents where there is a persistent tendency for the system to transition through the states $\text{T}\rightarrow \text{C}\rightarrow \text{P}$ rather than the reverse (Fig.\ \ref{fig:Fig4_EPR}(b)).

The foregoing observations prompt us to derive a corresponding theoretical model for the particle dynamics. Across the range of $\Gamma$ considered in our experiments, we find that the particles exhibit an exponentially decaying velocity-correlation function (VCF) and a mean-squared displacement (MSD) that is ballistic at short times and plateaus at long times due to confinement (Supplementary Section 2.2.5, Supplementary Fig.\ 6). These features of the particle motion can in principle be captured by a Langevin equation with a conservative force, however this corresponds to equilibrium dynamics. Further, while we do not observe significant inertial rebound of the particles--a hallmark of inertial active systems \cite{lowen2020inertial}--we do observe persistence in the particle trajectories arising from the coupling of the particles to the fluid-mediated memory of the non-equilibrium bath. This effect is known to be most pronounced when the particle radius is smaller than the Faraday wavelength \cite{xia2019tunable}, as in the present system. To encompass all of the foregoing behaviour, we project the particle dynamics onto a many-body active Ornstein-Uhlenbeck (AOU) model \cite{fodor2016howfar}, namely} 
\begin{subequations}
\label{eqn:AOUP}
\begin{align}
\label{eqn:AOUPa}
\mu\dot{\mathbf{x}}_i &= \sum_{j\neq i}\mathbf{F}_{ij} +\mathbf{F}_{\text{corral}} + \mu\mathbf{u}_i,\\
\label{eqn:AOUPb}
\tau\dot{\mathbf{u}}_i + \mathbf{u}_i &= \sqrt{2 D}\boldsymbol{\eta}_i,
\end{align}
\end{subequations}
where $\mu = 42.2\pm 2.9$ mg/s (1 mg = $1\times 10^{-3}$ g) is an experimentally determined drag coefficient (Supplementary Section 2.2.4 and Supplementary Fig.\ 5). The forces (whose analytical forms can be found in Supplementary Sections 2.2.1--2.2.5) on the right-hand side of Eqns.\ \eqref{eqn:AOUPa} from left-to-right are: a conservative, pairwise inter-particle force that combines the attractive capillary force and a power-law repulsive force to maintain particle separation upon contact; a confining force that mimics the square-shaped corral; and an active force where the propulsion velocities $\mathbf{u}_i$ follow an Ornstein-Uhlenbeck process with correlation time $\tau$ and diffusivity $D$ (Eqn.\ \eqref{eqn:AOUPb}). \sjt{In contrast to Brownian motion where $\tau \equiv 0$, the AOU model is out of equilibrium since finite-time correlations of the active force are not balanced by corresponding temporal correlations in the drag, violating the second fluctuation–dissipation relation \cite{kubo1966fluctuation}. Multi-particle interactions and the presence of the confining potential preclude writing down tractable analytical expressions for the VCF and MSD \cite{nguyen2021active, caprini2021inertial}. We therefore parameterize the model by simulating Eqns.\ \eqref{eqn:AOUP} and choosing the $D$ and $\tau$ that best fit experimental measurements of the VCF and MSD in the $L^2$-norm (Supplementary Section 2.2.5 and Supplementary Fig.\ 6).} Eqns.\ \eqref{eqn:AOUP} are then solved numerically using the semi-implicit Euler-Maruyama method to resolve stiffness arising through the contact forces (Methods, Supplementary Section 2.3) \cite{higham2002strong}. 

\sjt{Fit to the fundamental statistical observables of our system, the predictions of our theoretical model demonstrate excellent agreement with experiment. Both the transition rates (number of transitions per unit time) and resultant transition probabilities between microstates closely align with experimental values (Figs.\ \ref{fig:Fig3}(b) and (d)). Likewise, computing the steady-state occupation probabilities shows an analogous decrease in C states with increasing forcing and the accompanying increase in both T and P states (Fig.\ \ref{fig:Fig3}(e)). Most notably, the theoretical prediction of the EPR exhibits striking agreement with experiments across the range of $\Gamma$ (Fig.\ \ref{fig:Fig4_EPR}(a)) with broken detailed balance brought on by an analogous imbalance in cyclic probability currents in state space. We note that in our simulations when $\tau = 0$ we find no net circulation and $\sigma = 0$ as expected at equilibrium. Beyond quantitative agreement, the success of the theoretical model clarifies the fundamental physical processes underpinning the dynamics and emergent statistics in driven capillary self-assembly, specifically fluid-mediated memory and nonequilibrium driving of the fluid bath.}
\sjt{Beyond parameterizing the model for the case of six-particle clusters, it is natural to ask whether Eqns.\ \eqref{eqn:AOUP} have predictive capacity to capture the behaviour of larger assemblies. We therefore consider a seven-particle system driven at $\Gamma = 1.2g$ where, following the nomenclature of \cite{lim2019cluster}, the four microstates of the system are categorized as tree (Tr), turtle (Tu), flower (Fl), and boat (Bo) whose interaction energies are $U = -53.8, -53.2, -56.7$, and $-52.1$ nJ, respectively, rendering Fl the ground state. Compared to the three-state network of six-particle clusters, the four-state network of seven particles already implies a more complex set of possible transition pathways. Our simulations use the same parameter values for $D$ and $\tau$ as those calculated for the six-particle system at $\Gamma = 1.2g$. Like our six-particle scenario, there is excellent agreement between the experimental and theoretical predictions of the stationary distribution where we observe the ordering $\Pi(\text{Tr}) > \Pi(\text{Tu}) > \Pi(\text{Fl}) > \Pi(\text{Bo})$ (Fig.\ \ref{fig:Fig4_EPR}(c)). We note that these statistics are more akin to those expected at equilibrium in the high-temperature limit (Supplementary Fig.\ 1), yet in both experiment and theory we still observe a breakdown of detailed balance brought on by a corresponding imbalance in cyclic probability currents with an EPR of $\sigma = 0.0436\pm 0.0243\ \text{s}^{-1}$ and $\sigma = 0.0252\ \text{s}^{-1}$ in experiment and theory, respectively. The persistence of broken detailed balance despite near-equilibrium stationary distributions can be attributed to the richer cycle topology of the seven-particle network, which admits multiple independent cycles that provide additional pathways for nonequilibrium circulation (inset of Fig.\ \ref{fig:Fig4_EPR}(c)).}

\sjt{We have explored the emergent assembly statistics of millimetric particles bound by capillary forces at the fluid interface and driven out of equilibrium by supercritical Faraday waves. By directly measuring the entropy production rate both experimentally and theoretically, we demonstrate a clear breakdown of detailed balance in this macroscopic system. These findings highlight how nonequilibrium external driving biases transitions between metastable states that manifest as persistent probability currents in configuration space absent at equilibrium. The active Ornstein-Uhlenbeck model captures the essential physics of the system and points to the possibility that analogous phenomena may be observed and characterized in a broader class of active particle systems and athermal environments, including active bacterial baths \cite{angelani2011effective,ye2020active,grober2023unconventional, dhar2024active} or systems where the particles themselves are living or active \cite{buttinoni2012active,ko2022small,zottl2023modeling}. 

Our findings invite new strategies for controlling colloidal assembly \cite{perry2015two, liu2018capillary} in the presence of both thermal and active fluctuations. At this scale, the depth of the interaction potential at closest approach for a pair of particles is typically of the same order as the thermal energy. Hence thermal fluctuations can alone be sufficient to drive structural rearrangements between microstates and the stationary distribution is expected to follow equilibrium statistics. However, the addition of tunable active fluctuations (induced by light \cite{palacci2014light, rey2023light}, for example) opens the possibility of systematic exploration of otherwise inaccessible regions of the energy landscape. Coupled with engineered interparticle potentials \cite{gil1997statistical,lu2013colloidal}, this approach could enable rich programmable assembly pathways and the design of microstate distributions \cite{goodrich2021designing}. More broadly, our results reinforce the notion that macroscopic active matter systems can serve as versatile testbeds for general principles of non-equilibrium physics and stochastic thermodynamics, while theoretical modelling bridges insight into self-organization across disparate physical systems from microbial communities to robotic swarms and synthetic active materials.} 

This work is partially supported by the National Science Foundation (NSF CBET-2338320), the Office of Naval Research (ONR N00014-21-1-2816) and the Engineering and Physical Sciences Research Council (R104829). J.-W.B. is supported by the Department of Defense through the National Defense Science and Engineering Graduate (NDSEG) Fellowship Program. This work was carried out using the computational facilities of the Advanced Computing Research Centre, University of Bristol (http://www.bristol.ac.uk/acrc/). The data used in Figs.\ \ref{fig:Fig3} and \ref{fig:Fig4_EPR} are available from the corresponding author upon reasonable request.
\bibliography{nonEqCapillary}

\section{Methods}
The Supplementary Information provides full details of the experimental protocol (Supplementary Section 1), theoretical modelling, and simulation methods (Supplementary Section 2). 

\subsection{Experiments}
The spheres used in the experiment are commercially available (McMaster-Carr) PTFE spheres of diameter 0.159 centimeters. To reduce contact-line variability and hysteresis, the spheres are coated with a two-part hydrophobic coating (NeverWet Multisurface, Rust-Oleum). Four sets of six-sphere clusters are placed in one of four quadrants of a liquid bath. The bath is laser cut from acrylic and then filled with $9\pm 0.1$ milliliters of a water-glycerol solution. The entire set-up is mounted on a rigid aluminium platform and vibrated vertically using an electrodynamic modal shaker (Modal Shop, Model 2025E). To minimize the influence of ambient vibrations, the shaker is secured to a vibration-isolation table (ThorLabs Science Desk SDA75120). 

The spheres are filmed from above by a camera mounted directly over the fluid surface recording at 40 frames-per-second. The sphere locations are tracked during post-processing in MATLAB using the \texttt{imfindcircles} function and sphere trajectories are linked by computing the Euclidean distance between spheres in successive frames and choosing the sphere that minimizes this distance. The cluster configurations in each frame are identified from the clusters adjacency matrix which in turn allows us to compute the transition statistics between states over the lifetime of the experiment. 

\subsection{Theoretical modelling and numerical methods}
The active Ornstein-Uhlenbeck model \eqref{eqn:AOUP} contains three parameters, namely $\mu$, $D$, and $\tau$, that we fit to data from experiments. To determine the drag parameter $\mu$, we place two spheres on the fluid surface in the absence of Faraday waves and allow them to freely attract under the action of capillary forces. We compute a corresponding numerical simulation of Equations \eqref{eqn:AOUP} with $D = \tau = 0$ and choose the $\mu$ that best fits (in the $L^2$ norm) the Euclidean distance between the spheres as a function of time. The parameter $\mu$ remains fixed thereafter. The diffusion coefficient $D$ and correlation time $\tau$ are fit to the mean-squared displacement (MSD) and velocity correlation function (VCF) of the clusters at each value of the vertical acceleration $\Gamma$. Specifically, we perform a brute-force sweep over $D = (0.3, 0.31,\cdots, 1.8)$ and $\tau = (0.01, 0.02, \cdots, 0.3)$ and choose the $D$ and $\tau$ that best fit the MSD and VCF in the $L^2$ norm. 

Equations \eqref{eqn:AOUP} are a 24-dimensional system of stochastic differential equations that are solved using a semi-implicit Euler-Maruyama scheme in Python. The drift terms (the inter-particle contact force and the corral force) are updated implicitly while the stochastic terms are updated explicitly. At each time step of $\text{d}t = 1.25\times 10^{-4}$ s, the new particle locations are determined using the nonlinear root-finding algorithm \texttt{scipy.optimize.root}.

\end{document}